\documentclass[reprint, superscriptaddress, amsmath, amssymb, aps, pra]{revtex4-1}
\usepackage{graphicx}
\usepackage{sidecap}
\usepackage{dcolumn}
\usepackage{bm}
\usepackage[dvipdfm,colorlinks,linkcolor=blue, urlcolor=blue, anchorcolor=blue, citecolor=blue]{hyperref}
\usepackage{color}
\usepackage{ulem}
\usepackage{amsmath,amsfonts,amssymb}
\usepackage{booktabs}
\usepackage{float}
\usepackage{graphicx}

\begin{document}
\title{Quantum state engineering in a five-state chainwise system by coincident pulses technique}

\author{Jiahui Zhang}%
\email{y10220159@mail.ecust.edu.cn}
\affiliation{School of Physics, East China University of Science and Technology, Shanghai 200237, China}

\begin{abstract}
In this paper, an exact analytical solution is presented for achieving coherent population transfer and creating arbitrary coherent superposition states in a five-state chainwise system by a train of coincident pulses. We show that the solution of a five-state chainwise system can be reduced to an equivalent three-state $\Lambda$-type one with the simplest resonant coupling under the assumption of adiabatic elimination (AE) together with a requirement of the relation among the four incident pulses. In this method, all of four incident pulses at each step all have the same time dependence, but with specific magnitudes. The results show that, by using a train of appropriately coincident incident pulses, this technique enables complete population transfer, as well as the creation of arbitrary desired coherent superposition between initial and final states, while the population in all intermediate states is effectively suppressed. The complete physical explanation of the underlying
mechanism is presented. The results are of potential interest in applications where high-fidelity multi-state quantum control is essential, e.g., quantum information, atom optics, formation of ultracold molecules, cavity QED, nuclear coherent population transfer, light transfer in waveguide arrays, etc.
\end{abstract}
\maketitle

\section{\label{sec:level1}Introduction}
Coherent population transfer and coherent creation of superposition states attracted a great amount of attention for the control of chemical processes, quantum information processing, and nonlinear optics \cite{doi:10.1146/annurev.physchem.52.1.763, RevModPhys.79.53}. Potentially
scheme to produce them, based on he well-known stimulated Raman adiabatic passage (STIRAP) and its variants,
have been proposed \cite{RevModPhys.70.1003, 10.1063/1.4916903, RevModPhys.89.015006, PhysRevA.108.043710}. The standard STIRAP process utilizes a Raman transition with two counter-intuitively ordered laser pulses, allowing for the efficient and robust transfer of the population between an initial and a final state in a three-state system. The successes of STIRAP relies on the existence of the dark state, which should be followed adiabatically. The middle state, which is subjected to population decay in many physical implementations \cite{PhysRevA.56.1463, Ospelkaus2008, Li:17, Zhang_2021}, is not populated during the process because it is not present in the dark state.

To date, most investigations related with STIRAP focus upon the three-state $\Lambda$-type system, but STIRAP in atomic, molecular, and optical systems with four or more states is attracting increasing theoretical and experimental attention \cite{RevModPhys.70.1003, 10.1063/1.4916903, RevModPhys.89.015006, PhysRevA.108.043710}.
The success of STIRAP has prompted its extension to chainwise-connected multi-state systems \cite{SOLA2018151, PhysRevLett.109.260502, PhysRevA.45.4888, PhysRevA.48.845, PhysRevA.56.4929, PhysRevA.58.2295, PhysRevA.75.063832, PhysRevA.91.023802, PhysRevA.91.023802}, in which each state is connected only to its two neighbors: $|1\rangle\leftrightarrow|2\rangle\leftrightarrow|3\rangle\leftrightarrow\cdot\cdot\cdot\leftrightarrow|n\rangle$. In general, the goal is to transfer the population between the two ends without filling the intermediate states. The potential applications include many areas, such as: (\uppercase{\romannumeral1}) atomic mirrors and beams splitters in atom optics \cite{PhysRevLett.72.997}; (\uppercase{\romannumeral2}) atomic clocks \cite{PhysRevA.71.053410}; (\uppercase{\romannumeral3}) cavity QED \cite{PhysRevLett.71.3095, PhysRevA.51.1578}; (\uppercase{\romannumeral4}) spin-wave transfer via adiabatic passage in a five-level system \cite{Simon2007}; (\uppercase{\romannumeral5}) creation and detection
of ultracold molecules \cite{Danzl2010, Mark2009, PhysRevA.78.021402, PhysRevA.82.011609, PhysRevA.100.033421, PhysRevA.109.023109}, chainwise-STIRAP in M-type molecular system has been demonstrated a good alternative in creating ultracold deeply-bound molecules when the typical STIRAP in $\Lambda$-type system does not work due to weak Frank-Condon factors between the molecular states that are involved \cite{Danzl2010, PhysRevA.78.021402}; (\uppercase{\romannumeral6}) five-state nuclear coherent population transfer \cite{Amiri_2023}, which can be used in the study of the nucleus \cite{PhysRevC.87.054609}, the construction of nuclear batteries \cite{LIAO2011134, PhysRevC.105.064313}, as well as the construction of nuclear clocks that are much more accurate than atomic clocks \cite{Seiferle2019}; (\uppercase{\romannumeral7}) light transfer in waveguide arrays \cite{10.1063/1.2828985, PhysRevA.87.013806, PhysRevA.88.013808} and acoustic adiabatic propagation in multi-cavity chain system \cite{PhysRevApplied.14.014043}, which have profound impacts on exploring quantum technologies for promoting advanced optical and acoustic devices, and provides a direct visualization in space of typical phenomena in time. Quite often, it is necessary to achieve state preparation or transfer with high fidelity. This requires large temporal areas of the driving pulsed fields in order to suppress the non-adiabatic effects and thereby to make adiabatic evolution possible, which is, however, very hard to reach experimentally.

Although strategies for optimization of multi-state chainwise system have been developed, these come at the expense of strict relations on the pulse shapes \cite{PhysRevA.102.023515, 10.1063/5.0183063}, or require specific time-dependent nonzero detunings \cite{PhysRevA.100.033421}, or require some transient populations of the intermediate odd states \cite{PhysRevA.78.021402}, or even require additional couplings \cite{PhysRevA.102.023515}. Therefore, exploring alternative methods that can avoid the above problems is an important issue in the field of quantum control.
Remarkably, Rangelov and Vitanov have proposed a technique to complete population transfer in three-state systems by a train of $N$ pairs of coincident pulses \cite{PhysRevA.85.043407}, in which the population in the intermediate excited state is suppressed to negligible small value by increasing the pulse pairs. In this technique the number of pulse pairs is arbitrary and the robustness of system against deviation from exact pulse areas and spontaneous emission from excited state rise with increasing numbers of pulse pairs \cite{PhysRevC.94.054601}. Since the technique uses fields on exact resonance, the pulse shape is not important.
Recently, the technique has recently been generalized to tripod system \cite{PhysRevC.96.044619}. However, the application of coincident pulse technique in multi-state chainwise-connected systems has not been reported to date.

In this paper, an exact analytic solution is presented for achieving coherent population transfer and creating arbitrary coherent superposition states in a five-state chainwise system by a train of coincident pulses. Here we call the scheme five-state coincident pulses technique. The essence of this technique is to employ the reduced three-state propagator to achieve the desired population transfer by working in the parameter regime of exact two-photon resonances and large single-photon detunings as well as specific pulse sequences.
The results show that, by using $N (N\gg1)$ pairs of appropriately coincident incident pulses, this technique enables complete population transfer, as well as the creation of arbitrary desired coherent superposition between initial and final states with negligibly small transient populations in all the intermediate states. These properties make this technique an interesting alternative of the existing techniques for coherent control of five-state chainwise system.

\section{\label{sec:level2}Model and Method}
\begin{figure}[t]
\centering{\includegraphics[width=6cm]{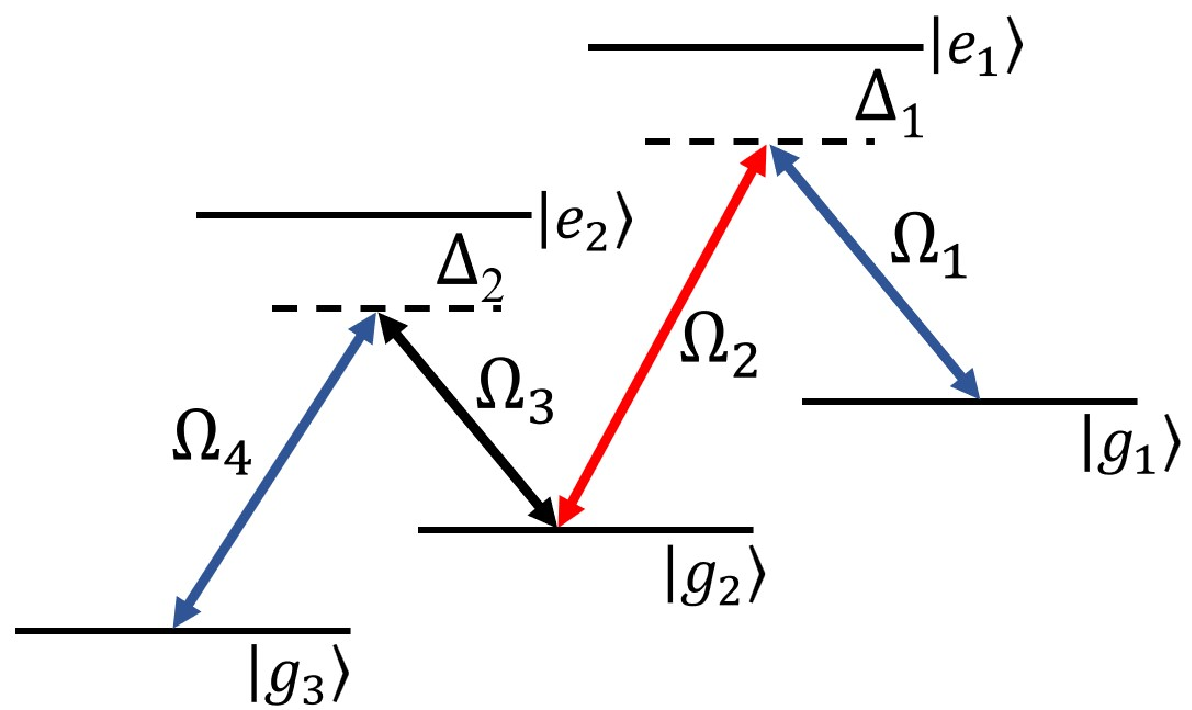}}
\caption{Linkage pattern of the chainwise-like five-state system.}
\label{fig1}
\end{figure}

The idea of this work can be described using a simple five-level system with states chainwise coupled by optical fields as illustrated in Fig.~\ref{fig1}. The states $| g_1\rangle$, $| g_2\rangle$ and $| g_3\rangle$ are three ground states while the intermediate states $| e_1\rangle$, and $| e_2\rangle$ refer to two excited states. The coupling between states is presented by the time-dependent Rabi frequency $\Omega_{i} (i=1, 2, 3, 4)$ in this figure.
The total wave function can be expanded as
\begin{eqnarray}\label{1}
| \psi(t)\rangle=c_{1}(t)| g_1\rangle&&+c_{2}(t)| e_1\rangle+c_{3}(t)| g_2\rangle\nonumber \\
&&+c_{4}(t)| e_2\rangle+c_{5}(t)| g_3\rangle.
\end{eqnarray}
the vector $c_{i}(t) (i=1, 2, 3, 4, 5)$ are the probability amplitudes of the corresponding state.
The evolution of system is described by the time-dependent Schr\"{o}dinger equation:
\begin{eqnarray}\label{2}
i\hbar\displaystyle\frac{\partial}{\partial t}c(t)=H(t)c(t).
\end{eqnarray}
Here, $c(t)$ is a five-component column matrix with the elements $\{c_{1}(t), c_{2}(t), c_{3}(t), c_{4}(t), c_{5}(t)\}$, and $|c_i(t)|^2$ stands for the corresponding probability.
In the interaction representation and after adopting the rotating-wave approximation, this system can be described in terms of a five-state Hamiltonian $(\hbar=1)$ \cite{RevModPhys.89.015006}
\begin{eqnarray}\label{3}
H(t)=\frac{1}{2}
\left(
\begin{array}{ccccc}
0&\Omega_{1}(t)&0&0&0\\
\Omega_{1}(t)&2\Delta_1&\Omega_{2}(t)&0&0\\
0&\Omega_{2}(t)&0&\Omega_{3}(t)&0\\
0&0&\Omega_{3}(t)&2\Delta_2&\Omega_{4}(t)\\
0&0&0&\Omega_{4}(t)&0\\
\end{array}
\right),
\end{eqnarray}
where the quantities $\Delta_{1}$ and $\Delta_{2}$ stand for one-photon detunings of the corresponding transitions.

If we assume that pairs of fields coupling two neighboring ground states are in a two-photon (Raman) resonance, the system has a dark state given by
\begin{equation} \label{4}
| \psi_0\rangle=\frac{\Omega_{2}(t)\Omega_{4}(t)| g_1\rangle-\Omega_{1}(t)\Omega_{4}(t)| g_2\rangle+\Omega_{1}(t)\Omega_{3}(t)| g_3\rangle}{\mathcal{N}(t)},
\end{equation}
where $\mathcal{N}(t)$ is a normalization factor. In ``conventional" five-state STIRAP, where the incident fields are applied in a counterintuitive time sequence. Adiabatically changing the Rabi frequencies of the optical fields so that the system stays in the dark state during evolution, one can transfer the system from the initial state $| g_1\rangle$ to the final state $| g_3\rangle$ without populating the short-lived intermediate state $| e_1\rangle$ and $| e_2\rangle$; however, the intermediate odd state $| g_2\rangle$ do acquire some transient populations along the way. In some particular systems these transient intermediate-state populations pose a problem because $| g_2\rangle$ may be a radiative state \cite{Simon2007, PhysRevA.78.021402}, decay from this state will degrade the coherent superposition (\ref{4}) and result in population loss from the dark state and reduction of the transfer efficiency; then it is necessary to reduce its population.

When the one-photon detunings are very large, meaning
\begin{eqnarray} \label{5}
|\Delta_1|&& \gg \sqrt{\Omega^2_{1}(t)+\Omega^2_{2}(t)},\nonumber\\
|\Delta_2|&& \gg \sqrt{\Omega^2_{3}(t)+\Omega^2_{4}(t)},
\end{eqnarray}
as we shall assume for our case \cite{PhysRevA.83.033830, PhysRevA.82.011609}, then states $| e_{1}\rangle$ and $| e_{2}\rangle$ are scarcely populated, and which can be adiabatically eliminated to obtain the following effective three-state Hamiltonian in the subspace \{$| g_1\rangle$, $| g_2\rangle$, $| g_3\rangle$\}:
\begin{eqnarray} \label{6}
H_e(t)
=\frac{1}{2}
\left(
\begin{array}{ccc}
\Delta_{e1}&\Omega_{e1}&0\\
\Omega_{e1}&\Delta_{e2}&\Omega_{e2}\\
0&\Omega_{e2}&\Delta_{e3}\\
\end{array}
\right),
\end{eqnarray}
where the two-photon Rabi frequencies are defined as
\begin{eqnarray}\label{7}
\Omega_{e_1}(t)&&=-\frac{\Omega_{1}(t)\Omega_{2}(t)}{2\Delta_1},\nonumber\\
\Omega_{e_2}(t)&&=-\frac{\Omega_{3}(t)\Omega_{4}(t)}{2\Delta_2}.
\end{eqnarray}
The three diagonal elements are
\begin{eqnarray}\label{8}
\Delta_{e_1}&&=-\frac{\Omega_{1}^2(t)}{2\Delta_1},\nonumber\\
\Delta_{e_2}&&=-\frac{\Omega_{2}^2(t)}{2\Delta_1}-\frac{\Omega_{3}^2(t)}{2\Delta_2},\nonumber\\
\Delta_{e_3}&&=-\frac{\Omega_{4}^2(t)}{2\Delta_2}.
\end{eqnarray}
Thereafter, an effective two-photon transition $|g_1\rangle\leftrightarrow|g_2\rangle$ and $|g_2\rangle\leftrightarrow|g_3\rangle$ will dominate. However, the system after AE subjects to dynamic Stark shifts from the trapping light, which can be expected to reduce the transfer efficiency.

If we assume that the three diagonal elements are equal to each other,
i.e., $\Delta_{e_1}=\Delta_{e_2}=\Delta_{e_3}=\Delta_{e}$.
This means that the four Rabi frequencies should satisfy
\begin{eqnarray}\label{9}
\Omega_{1}(t)=\sqrt{\zeta}\Omega_{4}(t)=\sqrt{\Omega_{2}^2(t)+\zeta\Omega_{3}^2(t)},
\end{eqnarray}
where $\zeta=\Delta_{1}/\Delta_{2}$.
By further setting $c_j =c^{'}_{j}e^{-i\Delta_{e} t}(j=1, 3, 5)$, the above equation can be written in the following form:
\begin{eqnarray}\label{10}
i\displaystyle\frac{\partial}{\partial t}
\left(
\begin{array}{ccc}
c^{'}_1\\
c^{'}_3\\
c^{'}_5\\
\end{array}
\right)
=\frac{1}{2}
\left(
\begin{array}{ccc}
0&\Omega_{e_1}(t)&0\\
\Omega_{e_1}(t)&0&\Omega_{e_2}(t)\\
0&\Omega_{e_2}(t)&0\\
\end{array}
\right)
\left(
\begin{array}{ccc}
c^{'}_1\\
c^{'}_3\\
c^{'}_5\\
\end{array}
\right),
\end{eqnarray} 
in which
\begin{eqnarray}\label{11}
\Omega_{e_1}(t)&&=-\frac{\Omega_{2}(t) \sqrt{\Omega_{2}^2(t)+\zeta\Omega_{3}^2(t)}}{2\Delta_1},\nonumber\\
\Omega_{e_2}(t)&&=-\frac{\Omega_{3}(t) \sqrt{\Omega_{3}^2(t)+\zeta^{-1}\Omega_{2}^2(t)}}{2\Delta_2}.
\end{eqnarray}
Consequently, a generalized $\Lambda$-type structure with the simplest resonant coupling  is formed. It is similar to STIRAP and allows for a complete STIRAP-like population transfer.

In order to implement the coincident pulse technique in this generalized model, we assume in the following that the effective Rabi frequencies $\Omega_{e_1}(t)$ and $\Omega_{e_2}(t)$ are pulse-shaped functions that share the same time dependence, but possibly with different magnitudes, 
i.e.,
\begin{eqnarray}\label{12}
\Omega_{e_1}(t)&&=\eta_1f(t),\nonumber\\
\Omega_{e_2}(t)&&=\eta_2f(t).
\end{eqnarray}
In this case, the Schr\"{o}dinger
equation (\ref{10}) is solved exactly by making a transformation to the so-called bright-dark basis \cite{PhysRevA.90.053837, PhysRevA.103.023527}.
The exact propagator is given by
\begin{widetext}
\begin{eqnarray}\label{13}
U(\varphi)=
\left(
\begin{array}{ccc}
1-2\sin^2\varphi\sin^2\frac{A}{4}&-i\sin\varphi\sin\frac{A}{2}&-2\sin2\varphi\sin^2\frac{A}{4}\\
-i\sin\varphi\sin\frac{A}{2}&\cos\frac{A}{2}&-i\cos\varphi\sin\frac{A}{2}\\
-2\sin2\varphi\sin^2\frac{A}{4}&-i\cos\varphi\sin\frac{A}{2}&1-2\cos^2\varphi\sin^2\frac{A}{4}\\
\end{array}
\right),
\end{eqnarray}
\end{widetext}
where $\tan\varphi=\Omega_{e_1}/\Omega_{e_2}=\eta_1/\eta_2$, the rms pulse area $A$ is defined
as $A=\int^t_{t_i}\sqrt{\Omega^2_{e_1}(t)+\Omega^2_{e_2}(t)}dt$.
According to propagator (\ref{13}), we can find the exact analytic solution for any initial condition.
For $\varphi=\pi/4$, which corresponds to $\eta_1=\eta_2$, and rms pulse area $A=2\pi$, the population of state $|g_1\rangle$ can be completely transferred to the final state $|g_3\rangle$.
However, Eqs.~(\ref{13}) implies that the intermediate ground state $|g_2\rangle$ will receive a significant transient populations.

In order to suppress the population of the intermediate ground state $|g_2\rangle$, one can use a sequence of $N$ pairs of
coincident pulse, each with rms pulse area $A(t)=2\pi$ at the
end of the $k$th step and mixing angles $\varphi_k$, the overall propagator is given by
\begin{eqnarray} \label{14}
U^{(N)}=U(\varphi_N)U(\varphi_{N-1})\cdot\cdot\cdot U(\varphi_{2})U(\varphi_{1}),
\end{eqnarray}
where $\varphi_k$ is given by
\begin{eqnarray}
\varphi_k&&=\frac{(2k-1)\pi}{4N}, k=1,2,...,N. \label{15}
\end{eqnarray}
The maximum population of the intermediate ground state $|g_2\rangle$ in the middle of each pulse pair is damped to small values by increasing the number of pulse.

Since the technique uses fields on exact resonance, the pulse shapes are unimportant. As an example, we assume the two-photon Rabi frequencies in the $k$th step are Gaussian shapes,
\begin{eqnarray}\label{16}
\tilde{\Omega}_{e_1}(t)&&=\Omega_{0}\sin\varphi_{k}e^{-(t-\tau_k)^2/T^2},\nonumber\\
\tilde{\Omega}_{e_2}(t)&&=\Omega_{0}\cos\varphi_{k}e^{-(t-\tau_k)^2/T^2},
\end{eqnarray}
where $\Omega_{0}= 2\sqrt{\pi}/T$ (corresponding to rms pulse area $A=2\pi$) and the mixing angles $\varphi_k (k=1,2,..., N)$ is given by Eq.~(\ref{15}).

Note that $\tilde{\Omega}_{e_1}(t)$ and $\tilde{\Omega}_{e_1}(t)$ are the two-photon Rabi frequencies for
transitions between $|g_1\rangle$ and $|g_2\rangle$ and between $|g_2\rangle$ and $|g_3\rangle$, which may be infeasible for many physical scenarios. Therefore, it is necessary to go back to the five-state
structure and design the physically feasible driving fields. Like Eqs.~(\ref{11}), we can impose~\cite{PhysRevA.94.063411}
\begin{eqnarray}\label{17}
\tilde{\Omega}_{e_1}(t)&&=-\frac{\tilde{\Omega}_{2}(t)\sqrt{\tilde{\Omega}_{2}^2(t)+\zeta\tilde{\Omega}_{3}^2(t)}}{2\Delta_1},\nonumber\\
\tilde{\Omega}_{e_2}(t)&&=-\frac{\tilde{\Omega}_{3}(t)\sqrt{\tilde{\Omega}_{3}^2(t)+\zeta^{-1}\tilde{\Omega}_{2}^2(t)}}{2\Delta_2}.
\end{eqnarray}

\noindent
Solving the two equations for $\tilde{\Omega}_{2}(t)$ and $\tilde{\Omega}_{3}(t)$ in Eq.~(\ref{17}), we
can express them as
\begin{eqnarray}\label{18}
\tilde{\Omega}_{2}(t)&&=
\tilde{\Omega}_a\sin\varphi_k e^{-(t-\tau_k)^2/\tilde{T}^2},\\
\tilde{\Omega}_{3}(t)&&=\tilde{\Omega}_b\cos\varphi_k e^{-(t-\tau_k)^2/\tilde{T}^2},
\end{eqnarray}

\noindent
where $\tilde{\Omega}_a=\sqrt{2\Delta_1\Omega_0}, \tilde{\Omega}_b=\sqrt{2\Delta_2\Omega_0},$ and $\tilde{T}=\sqrt{2}T$.

Accordingly, Rabi frequencies $\Omega_1(t)$ and $\Omega_4(t)$ are also modified following the similar relation as that in Eq.~(\ref{9}),
which is $\tilde{\Omega}_{1}(t)=\sqrt{\zeta}\tilde{\Omega}_{4}(t)=\sqrt{\tilde{\Omega}_{2}^2(t)+\zeta\tilde{\Omega}_{3}^2(t)}.$
Thus, we can obtain
\begin{eqnarray}
\tilde{\Omega}_{1}(t)&&=\tilde{\Omega}_ae^{-(t-\tau_k)^2/\tilde{T}^2},\\ \label{20}
\tilde{\Omega}_{4}(t)&&=\tilde{\Omega}_be^{-(t-\tau_k)^2/\tilde{T}^2}. \label{21}
\end{eqnarray}
According to expressions~(\ref{18})-(\ref{21}), it is not difficult to find that all four incident pulses at each step have the same time dependence, but have specific amplitudes, which is $\tilde{\Omega}_{1}(t):\tilde{\Omega}_{2}(t):\tilde{\Omega}_{3}(t):\tilde{\Omega}_{4}(t)=1:\sin\varphi_k:\sqrt{\zeta^{-1}}\cos\varphi_k:\sqrt{\zeta^{-1}}$,
and which can be simplified to $1:\sin\varphi_k:\cos\varphi_k:1$ if $\zeta=1$.
This pulse sequence is different from the previously proposed straddling-STIRAP~\cite{PhysRevA.56.4929} and alternating-STIRAP techniques~\cite{PhysRevA.44.7442}, which are two possible versions of STIRAP for multi-state systems with odd number of states.

Returning back to the five-state M-type system with chainwise coupling, the full Hamiltonian in Eq.~(\ref{3}) can be rewritten as
\begin{equation}\label{22}
H(t)=\frac{1}{2}
\left(
\begin{array}{ccccc}
0&\tilde{\Omega}_{1}(t)&0&0&0\\
\tilde{\Omega}_{1}(t)&2\Delta_1&\tilde{\Omega}_{2}(t)&0&0\\
0&\tilde{\Omega}_{2}(t)&0&\tilde{\Omega}_{3}(t)&0\\
0&0&\tilde{\Omega}_{3}(t)&2\Delta_2&\tilde{\Omega}_{4}(t)\\
0&0&0&\tilde{\Omega}_{4}(t)&0\\
\end{array}
\right).
\end{equation}
The above equation implies that we obtain a five-state coincident pulses
technique. With the modified Hamiltonian~(\ref{22}), this five-state chainwise system overcomes forbidden transitions among the ground states, while effectively suppressing all intermediate state populations in principle.

\section{\label{sec:level3}Results and Discussion}
In order to verify that the present protocol does work in the five-state chainwise system, we are going to employ Eq.~(\ref{2}) together with Eq.~(\ref{22}) to numerically investigate the population dynamics by using standard fourth-order Runge-Kutte method. For simplicity, we first focus on a closed system, assuming long coherence or absence of dephasing and dissipation.

\begin{figure}[H]
\centering{\includegraphics[width=8.7cm]{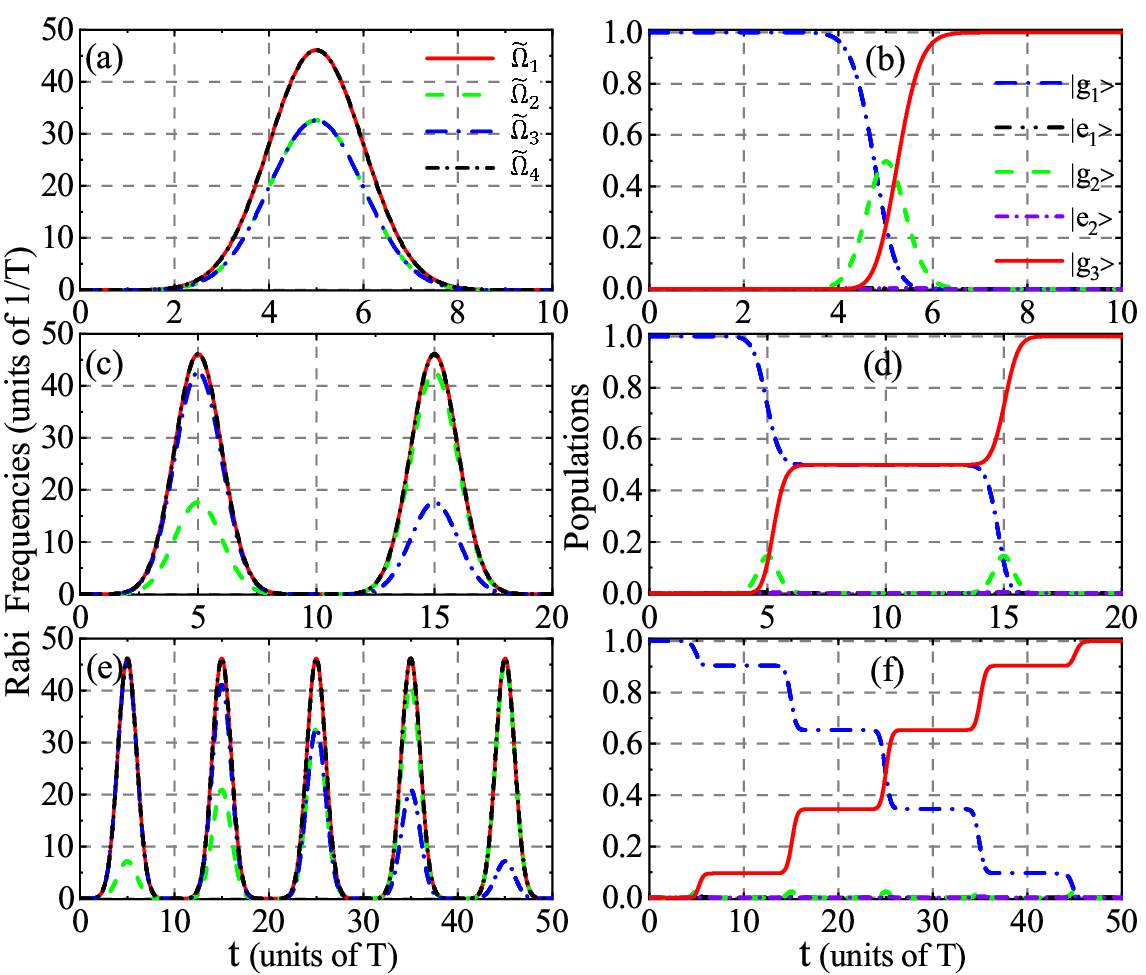}}
\caption{(Color online) Rabi frequencies (left column) and populations (left column) as a function of t for (from top to bottom) $N = 1, 2$, and $5$ pairs of pulses. Parameters used: $\Delta=300/T, \zeta=1$.}
\label{fig2}
\end{figure}
The left column of Fig.~\ref{fig2} show the corresponding Rabi frequencies when $\zeta=1$, which corresponds to the case of one-photon detunings $\Delta_1$ and $\Delta_2$ being equal. The right column of Fig.~\ref{fig2} show the corresponding population dynamics for several pulse trains of different number of pulse pairs. In all cases the population is transferred from state $|g_1\rangle$ to $|g_3\rangle$ in the end in a stepwise manner. As expected, the transient population of the intermediate state $|g_2\rangle$ is damped as $N (N=1, 2, 5)$ increases: from $0.5$ for a single pair of
pulses to about $0.024$ for $5$ pulse pairs. For this case, the analytic solution to maximum population in state $|g_2\rangle$ reads $P_{max}=\sin^2(\pi/4N)$. As $N$ tends to approach to infinity, the
maximum population of the middle state decreases as $1/N^2$. This suppression occurs on resonance and it results from the destructive interference of the successive interaction steps, rather than from large detuning or strong pulse. 
\begin{figure*}[t]
\centering{\includegraphics[width=18cm]{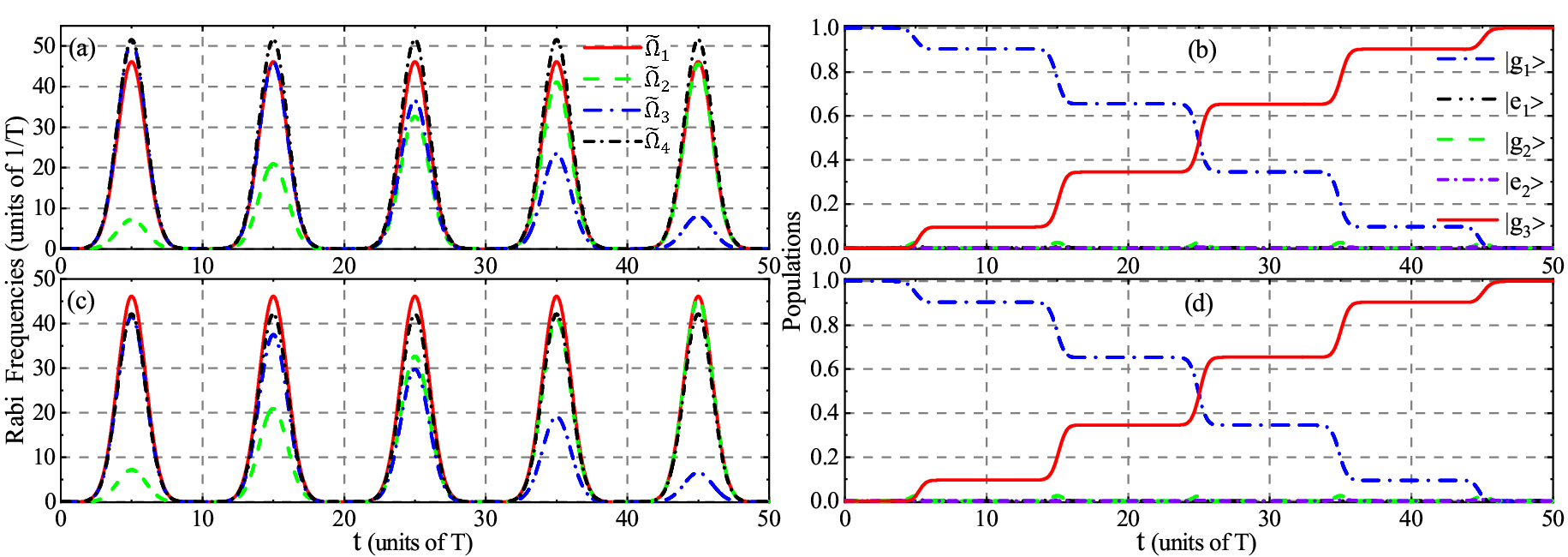}}
\caption{(Color online) Rabi frequencies (left column) and populations (left column) as a function of t for $N=5$ pairs of pulses. Relevant parameters are the same as in Fig.~\ref{fig2}(e) except (first row) $\zeta=0.8$ and (second row) $\zeta=1.2$.}
\label{fig3}
\end{figure*}
It is obvious that the excited states $|e_1\rangle$ and $|e_2\rangle$ are not involved during the transfer process, this is due to the AE protocol ensures the decoupling of the excited states from the dynamics so that we can directly move the population from state $|g_1\rangle$ to $|g_3\rangle$, $|e_1\rangle$ and $|e_2\rangle$ are only used to induce transitions but never significantly populated, as illustrated in Fig.~\ref{fig2}. Thus the transfer process is insensitive to the properties of excited states, e.g., fast spontaneous emission. This is very useful for depressing the effects of dissipation on the desired evolution of the system without relying on the dark state (\ref{4}).
Note that for $N\gg1$ the total pulse area is very large, it is useful when it is not possible to achieve high enough pulse areas needed for adiabatic evolution.

Below, let us limit the number of pulse pairs to $N=5$ for further demonstration. The left column of Fig.~\ref{fig3} show the corresponding Rabi frequencies when $\zeta=0.8$ and $1.2$ (from top to bottom), which corresponds to the cases of one-photon detunings $\Delta_1$ and $\Delta_2$ being slightly unequal. The right column of Fig.~\ref{fig3} show the corresponding population dynamics. Interestingly, we can see that these provide a comparable dynamic as when $\zeta=1$. 

\begin{figure}[H]
\centering{\includegraphics[width=8.7cm]{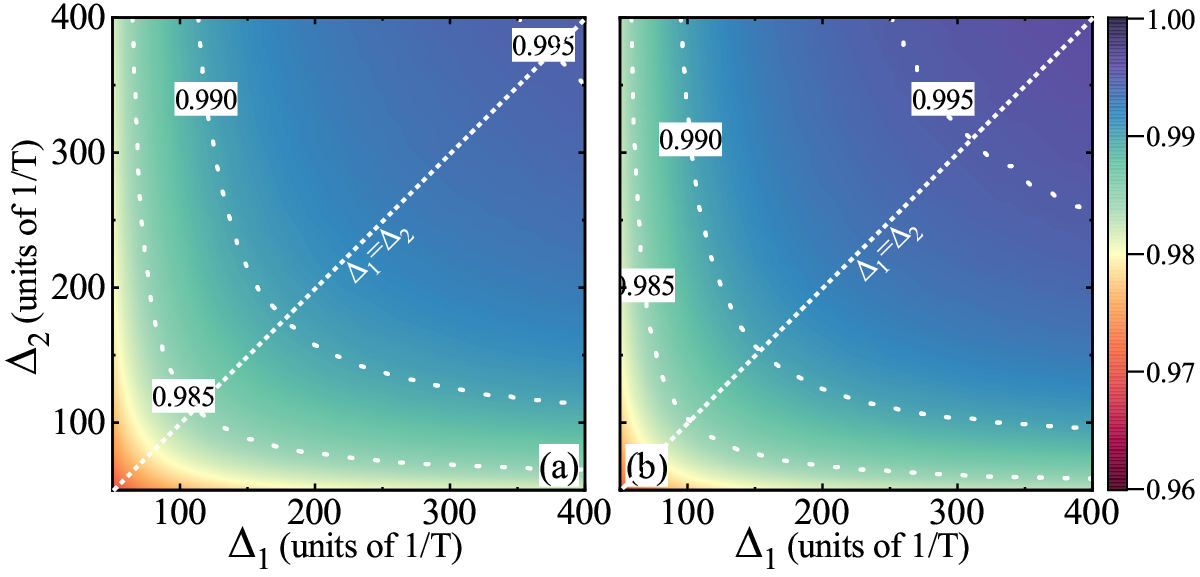}}
\caption{(Color online) Contour plot of the final population of state $|g_3\rangle$ against the one-photon detunings $\Delta_1$ and $\Delta_2$ for $N=5$ pairs.}
\label{fig4}
\end{figure}

In Fig.~\ref{fig4} we calculate the transfer efficiency as functions of two one-photon detunings
$\Delta_1$ and $\Delta_2$. In Fig.~\ref{fig4}(a), we consider the three intermediate states to be unstable and suffer from decay of rates $0.1/T$,  $0.01/T$ and $0.1/T$, while in Fig.~\ref{fig4}(b), we repeat the calculation of the transfer efficiency when the intermediate ground state $|g_2\rangle$ to be stable enough (in some particular system these transient intermediate-state populations do not pose a problem because these sublevels do not decay and there are no population losses \cite{PhysRevA.102.023515}). By comparison, we can clearly see that the transfer efficiency depends not only on the lifetime of $|g_2\rangle$ states but also on the lifetime of $|e_{1, 2}\rangle$ states. It can be expected that we can suppress the loss of $|g_2\rangle$ by adopting longer trains, while taking into account larger single-photon detunings to suppress the loss of excited states. In addition, as can seen from Fig.~\ref{fig4}, an efficient transfer efficiency requires that the value of $\Delta_1$ and $\Delta_2$ are roughly the same.

\begin{figure}[H]
\centering{\includegraphics[width=6.5cm]{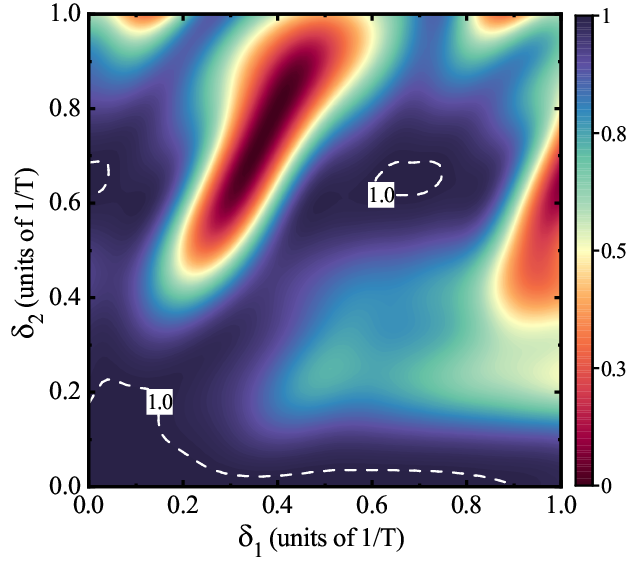}}
\caption{(Color online) Contour plot of the final population of state $|g_3\rangle$ against the two-photon detunings $\delta_1$ and $\delta_2$ for $N=5$ pairs. Relevant parameters are the same as in Fig.~\ref{fig2}(e).}
\label{fig5}
\end{figure}

In Fig.~\ref{fig5} we plot the final population versus the two-photon detuning, 
assuming in Eq.~(\ref{22}) that a field pair coupling two adjacent ground state is not strictly in two-photon Raman resonance, here $\delta_1$ and $\delta_2$ are the detunings of the first and second transitions. The figure demonstrates
that this technique is applicable for some range of non-zero
two-photon detunings.

\begin{figure*}[t]
\centering{\includegraphics[width=17cm]{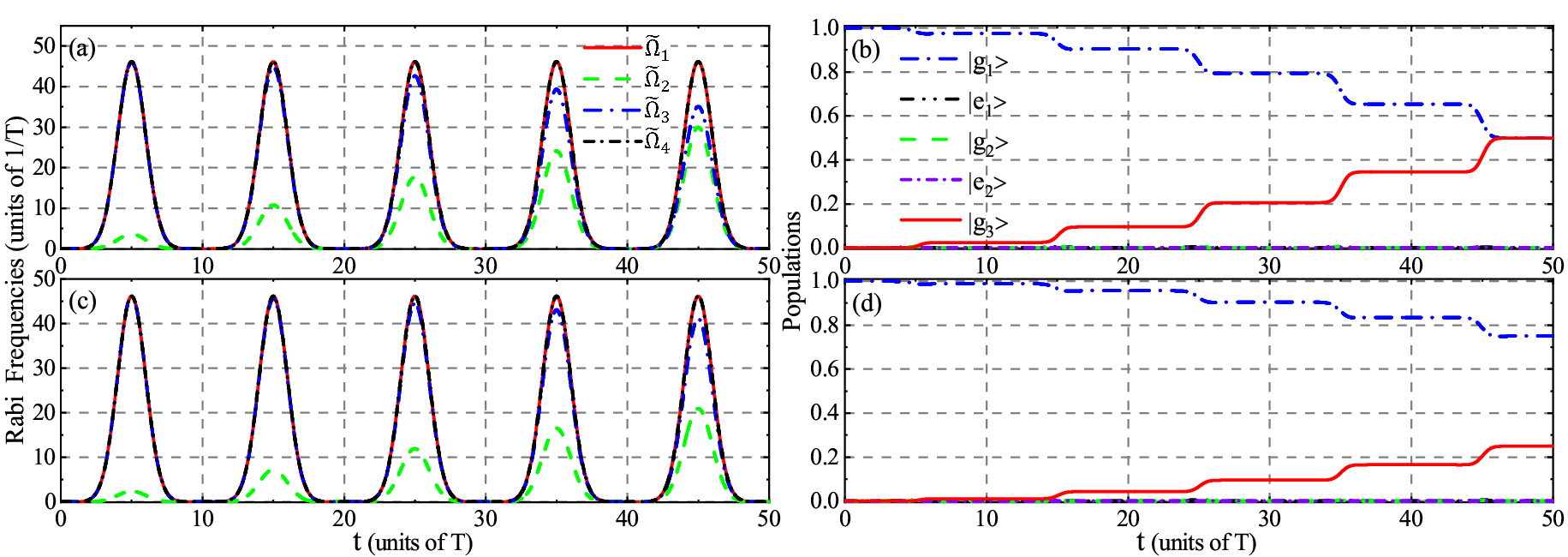}}
\caption{(Color online) Rabi frequencies (left column) and populations (left column) as a function of t for $N =5$ pairs of pulses. Relevant parameters are the same as in Fig.~\ref{fig2}(e) except (first row) $\varphi_k=[(2k-1)\pi]/(8N)$ and (second row) $\varphi_k=[(2k-1)\pi]/(12N)$.
}
\label{fig6}
\end{figure*}
More interestingly, we show that the five-state coincident pulses technique can achieve arbitrary superposition between $|g_1\rangle$ and $|g_3\rangle$ without populating almost all intermediate states. For example, $\varphi=\pi/8$ indicates a case of equal distribution according to Eq.~(\ref{13}). The corresponding Rabi frequency and population dynamics are shown in Fig.~\ref{fig6}(a) and Fig.~\ref{fig6}(b). For this case, the analytic solution to maximum population in state $|g_2\rangle$ reads $P_{max}=\sin^2(\pi/8N)$. As $N$ tends to approach to infinity, the
$P_{max}\approx(\pi/8N)^2$. The bottom portion of Fig.~\ref{fig6} shows a 3:1 example, in which $\varphi=\pi/12$. Of course, with reasonable modification of the value of $\varphi$,
arbitrary population distribution can also be expected.
\section{\label{sec:level5}Conclusion and outlook}
To conclude, we have generalized the standard coincident pulses technique to the five-state M-type chainwise system. The key of our protocol is reducing the M-type structure into a generalized $\Lambda$-type system with the
simplest resonant coupling, the simplification is realized under the assumption of AE together with a requirement of the relation among the four incident pulses. In this method, all of four incident pulses at each step all have the same time dependence, but with specific magnitudes. By using a train of $N (N\gg1)$ pairs of coincident incident pulses, this technique allows complete population transfer, as well as the creation of arbitrary coherent superpositions of the initial and final states without significant population of the three intermediate states.

A few other issues is worth mentioning. (\uppercase{\romannumeral1}): Given that the M-type five-state chain structure can be reduced to an equivalent $\Lambda$-type structure with the simplest resonant coupling (just like STIRAP), it can in principle be optimized by different methods, e. g., ``optimal control theory" \cite{PhysRevA.77.063420}, ``composite pulses" \cite{PhysRevLett.106.233001}, and ``shortcut-to-adiabatic passage" \cite{RevModPhys.91.045001}; (\uppercase{\romannumeral2}): The mechanism of the population transfer in five-state chainwise system can be optimized by tuning to a dressed middle state \cite{Vitanov1998, 10.1063/1.4916903}, and the dynamic can be basically reduced to that of the three resonant states \cite{10.1063/1.2828985}. Thereafter, this reduced system can be directly compatible with coincident pulses technique to achieve desired population transfer and creation of superposition states; (\uppercase{\romannumeral3}): Due to the feature of the chainwise system, i.e., each state is connected only to its two neighbors, so the ordering of the energy levels is not important: the linkage pattern may appear as a simple ladder-type, or as the M-type or W-type \cite{VITANOV200155}.

Finally, we believe that this technique is an interesting alternative of the existing techniques for coherent control of five-state chainwise systems, which has potential applications in high-fidelity quantum control of multi-state chainwise systems, e.g., quantum information \cite{RevModPhys.79.53}, atom optics \cite{PhysRevLett.72.997}, atomic clocks \cite{PhysRevA.71.053410}, formation of ultra-cold molecules \cite{Danzl2010, PhysRevA.78.021402}, cavity QED \cite{PhysRevLett.71.3095, PhysRevA.51.1578}, nuclear coherent population transfer \cite{PhysRevC.94.054601, Amiri_2023}, light transfer in waveguide arrays  \cite{10.1063/1.2828985, PhysRevA.87.013806}, etc.

\bibliography{references}

\end{document}